\documentclass[12pt,aps,prd,showpacs,superscriptaddress,nofootinbib]{revtex4-2}
\usepackage{amsmath}
\usepackage{amssymb}
\usepackage{color,graphicx}
\usepackage{tabularx}
\newcommand{\D}{\mathrm{d}}

\begin{document}

\title{A holographic analysis of the pion}
\author{Jeffrey Forshaw}
\email{jeffrey.forshaw@manchester.ac.uk}
\affiliation{Department of Physics and Astronomy,
  University of Manchester,
  Manchester M13 9PL,
  United Kingdom}
\author{Ruben Sandapen}
\email{ruben.sandapen@acadiau.ca}
\affiliation{Department of Physics, Acadia University, Wolfville, Nova Scotia, B4P 2R6, Canada}

\date{\today}

\begin{abstract}
Inspired by light-front holography, we compute the pion mass, charge radius, decay constant, electromagnetic form factor and electromagnetic transition form factor. To do so, we model the longitudinal quark dynamics using potentials due to 't Hooft and to Li \& Vary. We find a longitudinal wavefunction that is rather more peaked about $x\sim1/2$ than in previous studies. We also explore the strong degeneracy between these two potentials and conclude by noting that one scenario that accords well with the data also maps onto an equation previously noted by Vegh that describes the dynamics of a four-segmented string in AdS$_3$.

\end{abstract}


\maketitle

\section{Introduction}
\label{Sec:Intro}

The pion occupies a special status in particle physics. In the quark model, it is a quark-antiquark meson and yet it is much lighter than all other mesons. In QCD, it is simultaneously a bound state of quarks and gluons and a pseudo-Goldstone boson of chiral symmetry breaking. Its physical properties thus offer a unique window into the intertwined phenomena of confinement and chiral symmetry breaking \cite{Horn:2016rip}. This elusive physics is encoded in the Gell-Mann-Oakes-Renner relation (GMOR)\cite{GMOR}:
\begin{equation}
	f_\pi^2 M_\pi^2 = 2 m_{q} |\langle \bar{q} q \rangle| + \mathcal{O}(m_{q}^2),
\label{GMOR}
\end{equation}
where $M_\pi$ is the pion mass, $f_\pi$ its decay constant, $m_q$ is the light-quark mass\footnote{We assume isospin symmetry.} and $\langle \bar{q} q \rangle$ the quark condensate. Eq.~\eqref{GMOR} is supported by lattice calculations \cite{Becirevic:2004qv} and predicts that, in the chiral limit, the pion mass vanishes as $M_\pi^2 \propto m_{q}$ while $f_\pi$ remains finite.

Light-front holography (for a review, see \cite{Brodsky:2014yha}) is a realization of Maldacena's AdS/CFT correspondence \cite{Maldacena:1997re} in the conformal limit of QCD where quark masses and quantum loops are ignored. Light-front holography predicts that the pion is massless. To generate the physical pion mass, nonzero quark masses need to be taken into account. As a first approximation, this can be achieved using a prescription by Brodsky and de T\'eramond \cite{Brodsky:2008pg} known as the Invariant Mass Ansatz (IMA) which is widely used in the phenomenology of light mesons \cite{Brodsky:2014yha}. In a previous letter \cite{Forshaw:2012im}, we used the IMA with the light-front holographic wavefunction to successfully predict diffractive $\rho$ electroproduction. However, the IMA predicts that \cite{Li:2021jqb} $M_\pi^2 \approx 2m_{q}^2 \ln(\kappa^2/m_{q}^2 -\gamma_E)$, where $\gamma_E$ is the Euler constant,  instead of Eq. \eqref{GMOR}.  For this reason, the IMA becomes questionable, at least for the pion. 

Recently, there have been attempts to go beyond the IMA by modelling the longitudinal dynamics in mesons \cite{deTeramond:2021yyi,Li:2021jqb,Ahmady:2022dfv,Weller:2021wog,Ahmady:2021lsh} using the 't Hooft (tH) \cite{tHooft:1974pnl} and Li-Vary (LV) \cite{Li:2015zda} equations\footnote{Alternative models for longitudinal dynamics are reviewed in \cite{Li:2022izo}. For longitudinal dynamics with chiral perturbation theory constraints or in a  graviton soft-wall model, see \cite{Lyubovitskij:2022rod} or \cite{Rinaldi:2022dyh}.} so that Eq.~\eqref{GMOR} is satisfied.  Each of these papers has a different focus with the parameters of a given model varying significantly between the papers. With the exception of Ref.~\cite{Ahmady:2022dfv}, none of these papers consider simultaneously the three precisely measured, non-perturbative pion observables\footnote{$M_\pi=139.57039 \pm 1.8 \times 10^{-4}$ MeV, $f_\pi= 130.2 \pm 1.7$ MeV, $r_\pi=0.659 \pm 0.004$ fm. \cite{Workman:2022ynf}}: $M_\pi$, $f_\pi$ and the charge radius, $r_\pi$. Ref.~\cite{Ahmady:2022dfv} does so, together with form factor data, but overestimates $f_\pi$. 

Among light mesons, the pion is most sensitive to longitudinal dynamics since, as we shall see, its mass is entirely generated by it. Our goal in this paper is to focus exclusively on the pion and in so doing we are able to explore the well-established degeneracy between the ground-states of the tH and LV models \cite{tHooft:1974pnl,deTeramond:2021yyi,Weller:2021wog}. We also point out that an equation due to Vegh \cite{Vegh:2023snc}, obtained by studying the dynamics of a segmented string in AdS$_3$, coincides with a viable scenario.

\section{Transverse and longitudinal dynamics}
Our starting point is light-front QCD, where the Schr\"odinger-like equation for mesons reads \cite{Brodsky:1997de} 
\begin{equation}
	\left(\frac{-\nabla^2}{x(1-x)} + \frac{m^2_q}{x}+ \frac{m^2_{\bar{q}}}{1-x} +	U(x,\mathbf{b}) \right) \Psi (x,\mathbf{b}) = M^2 \Psi(x,\mathbf{b}), 
\label{LFSE-meson}
\end{equation}
where $M$ is the meson mass and $\Psi(x,\mathbf{b})$ is the meson's wavefunction, with  $x$ being the light-front momentum fraction carried by the quark and $\mathbf{b}=(b_\perp, \varphi)$ the transverse displacement of the quark and antiquark. Assuming isospin symmetry, we take $m_u=m_d\equiv m_{q}$. The confining potential, $U$, cannot yet be derived from first principles in QCD although, as we shall see, progress has been achieved using light-front holography and its extensions. The  normalization condition on the wavefunction is
\begin{equation}
	\int \mathrm{d}^2 \mathbf{b} \, \mathrm{d} x \, |\Psi(x,\mathbf{b})|^2 = 1 ,
\label{normalization-b}
\end{equation}
which embodies the assumption that the pion consists only of the leading $q\bar{q}$ Fock sector.  

We introduce the transverse vector $\boldsymbol{\zeta}=\sqrt{x(1-x)} \mathbf{b}$ such that, under the assumption~\cite{Chabysheva:2012fe},
\begin{equation}
	U(x,\mathbf{b}) = U_\perp(\zeta) + U_\parallel(x) ,
\label{Uzetax}
\end{equation}
Eq. \eqref{LFSE-meson} can be solved by a separation of variables, i.e. 
\begin{equation}
	\Psi (x, \mathbf{b})= \frac{\phi (\zeta)}{\sqrt{2\pi \zeta}} e^{i L \varphi} X(x),	
\label{full-mesonwf}
\end{equation}
where $L$ is the relative orbital angular momentum of the $q\bar{q}$ pair.
Eq. \eqref{LFSE-meson} separates into 
\begin{equation}
	\left[-\frac{\D^2}{\D \zeta^2}+\frac{4L^2-1}{4 \zeta^2}+U_\perp \right] \phi(\zeta)= M_\perp^2 \phi(\zeta)
\label{hSE}
\end{equation}
and
\begin{equation}
	 \left[\frac{m^2_q}{x(1-x)} +U_\parallel \right] X(x) =M_\parallel^2 X(x) 
\label{parallel}
\end{equation}
with $M^2=M_\perp^2 + M_\parallel^2$. Choosing 
\begin{equation}
\int_0^\infty \mathrm{d} \zeta \, |\phi(\zeta)|^2=1\;,
\label{normalization-zeta}
\end{equation}
Eq. \eqref{normalization-b} implies that 
\begin{equation}
\int_0^1 \frac{\mathrm{d} x}{x(1-x)} |X(x)|^2 = 1 \;. 
\label{normalization-X}
\end{equation}
It is useful to define $\chi(x)=X(x)/\sqrt{x(1-x)}$, so that Eq. \eqref{parallel} becomes
\begin{equation}
	 \left[\frac{m_q^2}{x(1-x)} + V_\parallel \right] \chi(x) =M_\parallel^2 \chi(x) ,
\label{parallel-chi}
\end{equation}
where
\begin{equation}
  V_\parallel=  \frac{1}{\sqrt{x(1-x)}} U_\parallel \sqrt{x(1-x)} 
\label{V-U-parallel}
\end{equation}
and 
\begin{equation}
\int_0^1 \mathrm{d} x \, |\chi(x)|^2=1 \;. 
\label{normalization-chi}
\end{equation}
Interestingly, Eq.~\eqref{parallel-chi} coincides with the tH equation for mesons in  $(1+1)$-dimensional, large-$N_c$ QCD; suggesting that $V_\parallel$ may be the 't Hooft potential operator \cite{tHooft:1974pnl}. This is the choice made in Refs.~\cite{Ahmady:2021lsh,Ahmady:2022dfv}. On the other hand, Ref.~\cite{Chabysheva:2012fe}, which was the first paper to extend light-front holography by accounting for longitudinal dynamics, chooses $U_\parallel$ to be the tH potential operator. Alternatively, Refs.~\cite{deTeramond:2021yyi,Li:2021jqb} choose $V_\parallel$ to be the LV potential operator \cite{Li:2015zda}. 

 More generally, the eigenspectrum of Eq. \eqref{parallel-chi} is invariant under the similarity transformation
 \begin{equation}
    V_\parallel \to h(x) \, V_\parallel \, \frac{1}{h(x)}
\label{similarity-V}
\end{equation}
for some $h(x)$, while the eigenfunctions transform as
\begin{align}
    \chi(x) &\to h(x) \chi(x)~.
\label{similarity-chi}
\end{align}
As we will motivate later, we consider $h(x) = [x(1-x)]^{n/2}$ and take $V_\parallel$ to be the tH or LV operator as our default (i.e. $n=0$). In this language, Ref.~\cite{Chabysheva:2012fe} takes $n=-1$ while Refs.~\cite{deTeramond:2021yyi,Li:2021jqb,Ahmady:2022dfv,Weller:2021wog} choose $n=0$. It turns out that $n<0$ is badly behaved in the chiral limit (see Appendix \ref{App:chiral}).  We will also see that the pion data exclude $n\ge3$ while the remaining possibilities, $n=0,1,2$ are all able to fit the data. We refer to the $n=0,1,2$ cases as Models A, B, C respectively. For example, Model tH-B means that $V_\parallel$ is the 't Hooft potential operator with $n=1$.   

Eq. \eqref{hSE} is the holographic Schr\"odinger Equation, derived by Brodsky and de T\'eramond \cite{deTeramond:2008ht}, where the variable $\zeta$ maps onto the fifth dimension in $\mathrm{AdS}_5$, with \cite{Brodsky:2014yha}
\begin{equation}
	U_\perp = U_{\mathrm{LFH}}=\kappa^4 \zeta^2 + 2 \kappa^2 (J-1) \;,
	\label{holographic-potential}
\end{equation}
where $J$ is the spin of the meson and $\kappa$ is an emerging confinement scale. It is argued in \cite{Brodsky:2013ar} that the analytic form of $U_\perp$, given by Eq. \eqref{holographic-potential}, is uniquely fixed by the underlying conformal symmetry and holographic mapping to $\mathrm{AdS}_5$. The mass scale, $\kappa$,  in the harmonic (first) term of Eq. \eqref{holographic-potential} does not spoil the conformal symmetry of the underlying action. Then, the holographic mapping implies that a quadratic dilaton deforms the geometry of pure $\mathrm{AdS}_5$ (this is referred to as the soft wall model \cite{Karch:2006pv})  and this fixes the constant (second) term in Eq. \eqref{holographic-potential}. 

Solving Eq. \eqref{hSE} using Eq. \eqref{holographic-potential} yields
\begin{equation}
	M_{\perp}^2(n_\perp, J, L)= 4\kappa^2 \left( n_\perp + \frac{J+L}{2} \right) \;,
\label{MT}
\end{equation}
which predicts that the lowest lying meson, with $n_\perp=J=L=0$, is massless. It is natural to identify this state with the pion. The corresponding wavefunction is~\cite{Brodsky:2014yha} 
\begin{equation}
	\phi_{\pi}(\zeta) =\kappa \sqrt{2\zeta} \exp{\left(\frac{-\kappa^2 \zeta^2}{2}\right)} \;.
	\label{pionwf-gaussian}
\end{equation}
Using Eq.\eqref{full-mesonwf}, the pion wavefunction is thus
\begin{equation}
    \Psi_\pi(x,\mathbf{b})=\frac{\kappa}{\sqrt{\pi}} \sqrt{x(1-x)} \; \chi(x) \exp{\left(\frac{-\kappa^2 x(1-x) b_\perp^2}{2}\right)},
    \label{pionwf}
\end{equation}
where $\chi(x)$ is now the lowest eigenfunction of Eq.~\eqref{parallel-chi}. 

Armed with Eq. \eqref{pionwf} and a form for $V_\parallel$, we are able to compute a number of important properties of the pion. Specifically, we focus on the pion mass ($M_\pi$), decay constant ($f_\pi$),  electromagnetic form factor ($F_\pi(Q^2)$) and the pion-to-photon transition form factor ($F_{\pi \gamma} (Q^2)$).
The pion mass is given by Eq. \eqref{parallel-chi}: 
\begin{equation}
	M_\pi^2 = \int_0^1 \mathrm{d}x \; \chi^*(x) \left[\frac{m_q^2}{x(1-x)} + V_\parallel \right] \chi(x) 
	\label{Mpi-chi} \;
\end{equation}
and the other quantities can be computed using (e.g. see \cite{PhysRevD.102.036005,Ahmady:2021lsh}) 
\begin{align}
    f_\pi &= \sqrt{\frac{6}{\pi}} \int_0^1 \mathrm{d} x \;\Psi_\pi(x,\mathbf{0}) \nonumber \\ &= \frac{\sqrt{6}}{\pi} \kappa \int_0^1 \mathrm {d} x \sqrt{x(1-x)} \chi(x) \;,
\label{fpi}
\end{align}
\begin{align}
    F_\pi(Q^2) & = \pi \int_0^1 \mathrm{d} x \; \D b_\perp^2 \; J_0((1-x) b_\perp Q) \; |\Psi_\pi(x,\mathbf{b})|^2 \nonumber \\ & = \int_0^1 \mathrm{d} x \; |\chi(x)|^2 \; \exp \left(-\frac{(1-x)}{x} \frac{Q^2}{4\kappa^2} \right) 
\label{EMFF}
\end{align}
and
\begin{align}
\label{TFF}
	Q^2 F_{\pi \gamma} (Q^2)&=\frac{2 \kappa}{\sqrt{3} \pi}  \int_0^1 \mathrm{d} x \sqrt{x(1-x)} \chi(x) \\ \nonumber 
	\times & \int_0^\infty \mathrm{d} b_\perp (m_q b_\perp) K_1(m_q b_\perp) 
	 \exp\left(-\frac{\kappa^2 x(1-x) b_\perp^2}{2} \right) 
	Q J_1(b_\perp (1-x) Q) \;,
\end{align}
which is derived in Appendix A.

There are two further quantities of interest. The pion charge radius, which is related to the $Q^2 \to 0$ limit of the form factor (FF),
\begin{align}
    r^2_{\pi}&=  -6 \lim_{Q^2 \to 0}\frac{\mathrm{d}F_{\pi}(Q^2)}{\mathrm{d}Q^2} \nonumber \\ 
     & =  \frac{3}{2\kappa^2} \int_0^1 \mathrm{d} x \frac{(1-x)}{x} |\chi(x)|^2,
\label{rpi}
\end{align}
and the $\pi^0 \to \gamma \gamma$ decay width, which is related to the $Q^2 \to 0$ limit of the transition form factor (TFF): 
\begin{equation}
	\Gamma_{\gamma \gamma} = \frac{\pi}{4} \alpha_{\mathrm{em}}^2 M_\pi^3 |F_{\pi \gamma}(0)|^2 \; .
    \label{eq:gwidth}
\end{equation}
Our attention must now turn to the form of the longitudinal potential, $V_\parallel$, and its lowest-lying eigenfunction.

\section{The Li-Vary model} 
\label{Sec:LV}
A model for $V_\parallel$ that admits an exact power-law solution for its ground state is the phenomenological QCD potential first proposed by Li, Maris, Zhao and Vary \cite{Li:2015zda} and further studied by Li and Vary (LV) \cite{Li:2021jqb}. The LV potential operator is
\begin{equation}
	V_\parallel^\text{LV} = -\sigma^2 \partial_x (x(1-x) \partial_x)
\label{LV}
\end{equation} 
and the lowest-lying eigenfunction is \cite{Li:2021jqb,deTeramond:2021yyi} 
\begin{equation}
\chi(x) =  N(\beta)\; x^\beta (1-x)^\beta
\label{LV-soln}
\end{equation}
with $\beta=m_q/\sigma$. We see that $\chi(x) \to 1$ in the chiral limit ($m_q \to 0$), which is the same form as that predicted using holography in \cite{Brodsky:2006uqa,Brodsky:2007hb,Brodsky:2008pf}. The various observables are then given by
\begin{align}
M_\pi &= \sigma \sqrt{2 \beta (1+2 \beta)} \label{massLV} \\
    f_\pi &= N(\beta) \; \kappa \sqrt{\frac{6}{\pi}} \frac{1}{2^{2(1+\beta)}} \; \frac{\Gamma(3/2+\beta)}{\Gamma(2+4\beta)} ,
    \label{fpiLV} \\
    r_\pi^2 &= \frac{3(1+2\beta)}{4\kappa^2\beta}
\label{rpiLV} \\
    F_\pi(Q^2)&= N(\beta)^2 \; \Gamma(1+2 \beta) \;U\left(1+2\beta,-2\beta,\frac{Q^2}{4\kappa^2}\right) ,
\label{FpiLV}
\end{align}
where $U$ is the confluent hypergeometric function of the second kind and 
\begin{align}
    N(\beta)^2 &= \frac{\Gamma(2+4 \beta)}{\Gamma(1+2\beta)}. 
    \end{align}

For the potentials given by Eq. \eqref{similarity-V}, all observables except for Eq.~\eqref{massLV} can still be computed using the above equations with $n$ successive replacements $\beta \to \beta + 1/2$. The pion mass is the exception since it is invariant under the similarity transformations (recall that $M^2_\pi=M^2_\parallel$, i.e. it is the eigenvalue of Eq.~\eqref{parallel-chi}). Note that Eq. \eqref{massLV} implies that
\begin{equation}
    M_\pi^2 =2 \sigma m_q + 4 m_q^2,
    \label{LV-GMOR}
\end{equation}
i.e. the LV models are consistent with the GMOR relation.

\begin{table}
\centering
\begin{tabularx}{0.45\textwidth}{|>{\centering\arraybackslash}X|>{\centering\arraybackslash}X|>{\centering\arraybackslash}X|>{\centering\arraybackslash}X|}
\hline
         & $\kappa$/MeV & $m_q$/MeV & $\sigma$/MeV \\ \hline
    LV-A & 423 & 60.4 & 40.3\\ \hline
    LV-B & 423 & 57.0 & 56.9\\ \hline
    LV-C & 423 & 49.4 & 98.5 \\ \hline
\end{tabularx}
    \caption{Extracted values of the parameters of the LV model discussed in the text.}
    \label{tab:LV}
\end{table}

Since the decay constant and radius depend only upon $\kappa$ and $\beta$ we use their measured values to fix these two parameters. The remaining parameter, $\sigma$, can then be determined from the pion mass. In Table \ref{tab:LV} we show the resulting values of $\kappa$, $m_q$ and $\sigma$ for the LV-A, LV-B and LV-C models. Notice that the preferred values of $\beta = m_q/\sigma$ are very close to $1.5, 1.0$ and $0.5$. This means that $X(x) \propto [x(1-x)]^2$, which is considerably more peaked than the $X(x) \propto \sqrt{x(1-x)}$ form anticipated in \cite{Brodsky:2006uqa,Brodsky:2007hb,Brodsky:2008pf}.

\begin{figure}[ht]
\centering 
\includegraphics[width=12.5cm]{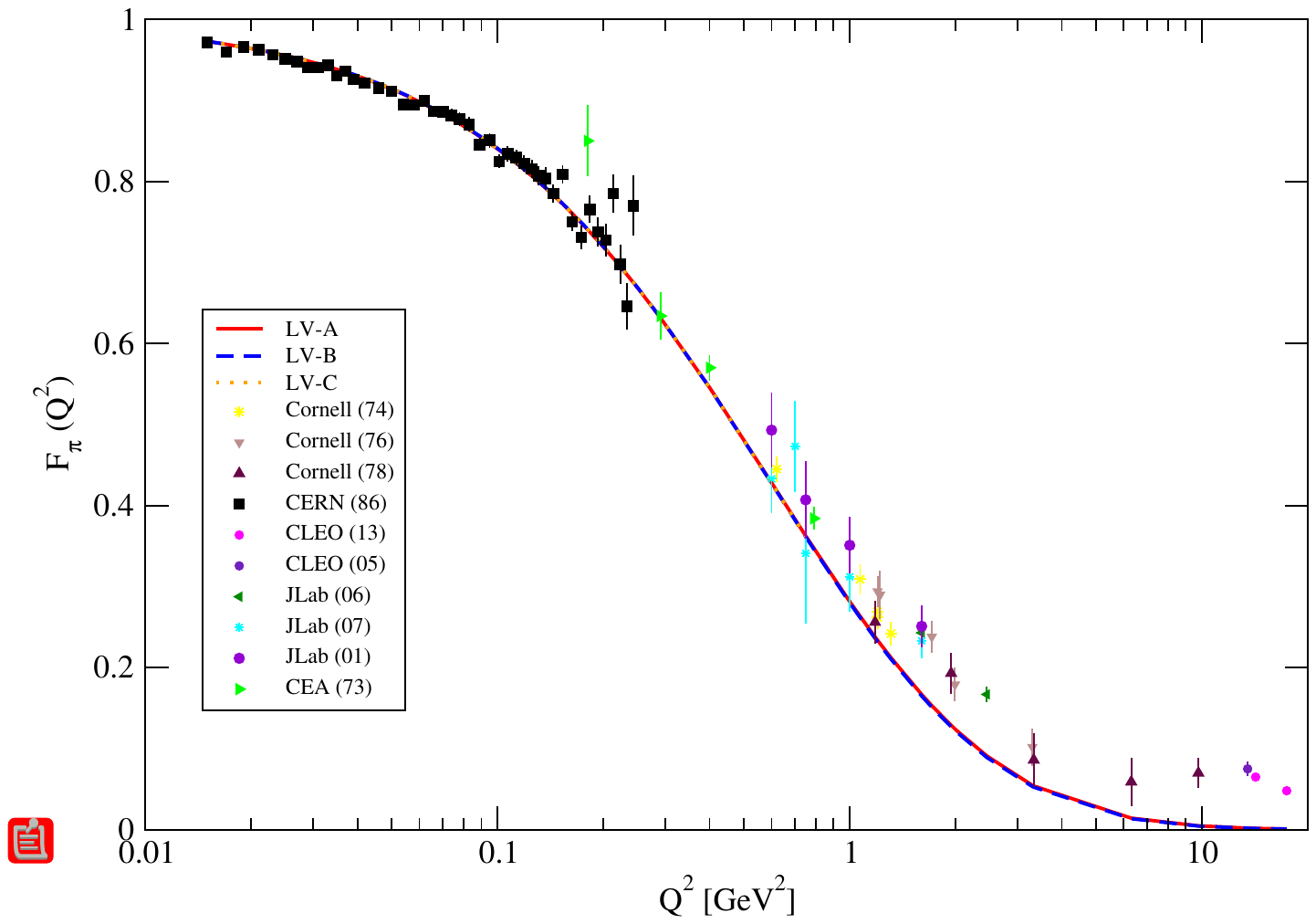}\\ \includegraphics[width=12.5cm]{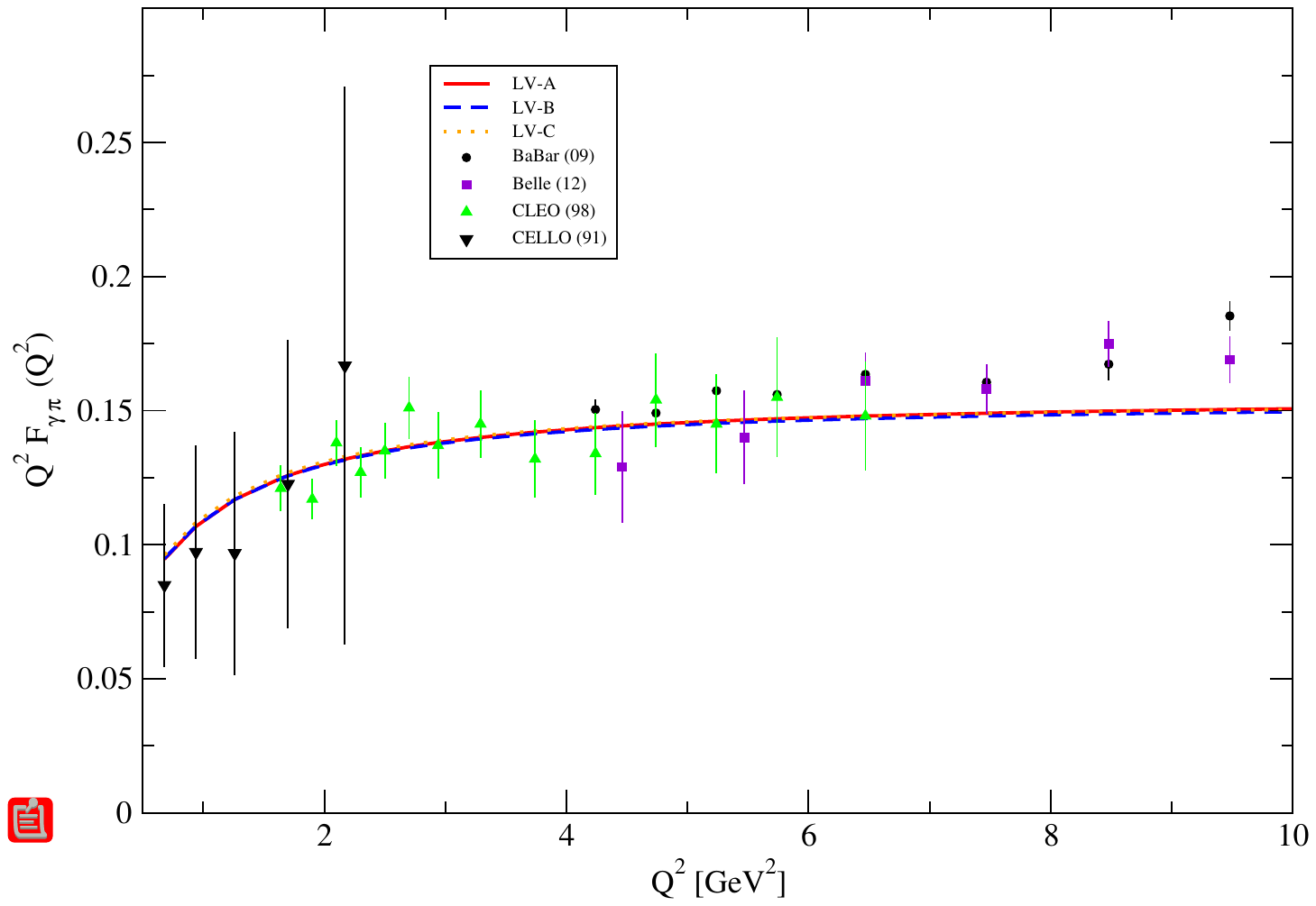}\caption{Postdictions of the LV models for the FF data \cite{Amendolia:1986wj,Bebek:1974,Bebek:1976,Bebek:1978,Volmer:2001,Horn:2006tm,Pedlar:2005sj,Seth:2012nn} and TFF data \cite{Uehara2012,Aubert2009,Gronberg1998, 
Behrend1991}.} 
\label{Fig:FF-LV}
\end{figure}

We are now in a position to postdict the form factor data and in Figure \ref{Fig:FF-LV} we show the results for the LV models. Agreement with the data is excellent for low values of $Q^2$, which accords with the fact that we do not include any perturbative QCD evolution. All three forms of the LV potential give identical results for the FF, since it depends only upon $\kappa$ and $\beta$. The TFF has some very weak dependence on $m_q$.

\clearpage

The decay width that we obtain using the $Q^2 \to 0$ limit of Eq.~\eqref{TFF} is $\Gamma_{\gamma\gamma} = 7.0, 7.2,7.4$~eV for LV-A, LV-B, LV-C. This compares favourably to the experimental measurement of $7.82~\pm~0.22$~eV \cite{Workman:2022ynf}, though we do note that Eq.~\eqref{TFF} is missing non-perturbative corrections to the photon wavefunction as $Q^2 \to 0$. Better agreement with experiment is obtained if we use the Adler-Bell-Jackiw (ABJ) chiral anomaly relation,
\begin{equation}
 	F_{\pi \gamma}(0) = \frac{1}{2\sqrt{2} \pi^2 f_\pi} \;,
 \label{ABJ-lit}
 \end{equation}
 in Eq.~\eqref{eq:gwidth}, in which case all three models necessarily give $\Gamma_{\gamma\gamma} = 7.8$~eV. This is simply the statement that the data are in accord with ABJ.

\section{The 't Hooft model}
\label{Sec:tH}
Another candidate for $V_\parallel$ is the 't Hooft model (tH), in which case Eq. \eqref{parallel-chi} becomes \cite{tHooft:1974pnl}
\begin{equation}
    \frac{m^2_q-g^2}{x(1-x)} \chi(x) - g^2 \mathcal{P} \int_0^1 {\rm d}y \frac{\chi(y)}{(x-y)^2} = M_\parallel^2 \chi(x),
\label{tH}
\end{equation}
where the principal value prescription is defined as
\begin{equation}
	\mathcal{P} \int \mathrm{d} y \frac{f(x,y)}{(x-y)^2} \equiv \lim_{\epsilon \to 0^{+}}\frac{1}{2}\int \mathrm{d} y \left(\frac{f(x,y)}{(x-y+i\epsilon)^2} + \frac{f(x,y)}{(x-y-i\epsilon)^2}\right) \;.
\end{equation}
No analytical solution to Eq.~\eqref{tH} is known and it has to be solved numerically. However, it is known that  $\chi(x) \sim x^\beta$ or $\chi(x) \sim (1-x)^\beta$ at the end-points, $x \to 0,1$,  as a consequence of the Hamiltonian being hermitian \cite{tHooft:1974pnl,Ambrosino:2023dik}. In fact, as illustrated in Fig.~\ref{Fig:chi}, $\chi(x) \sim [x(1-x)]^\beta$ is a very good approximation for the ground state wavefunction in the tH model across the entire range in $x$. The end-point analysis of Eq.~\eqref{tH} yields: 
\cite{tHooft:1974pnl,Chabysheva:2012fe}
\begin{equation}
    m_q^2 -g^2 + g^2 \pi \beta \cot (\pi \beta)  =0 \;,
\label{end-points-tH-A}
\end{equation}
where we have used  \cite{Ambrosino:2023dik}
\begin{equation}
 -\mathcal{P} \int_0^1 \mathrm{d} y \frac{\chi(y)}{(x-y)^2} \to -\mathcal{P} \int_0^\infty \mathrm{d} y \frac{y^\beta}{(x-y)^2}=\pi \beta \cot(\pi \beta) x^{\beta-1}
\end{equation}
if $x \to 0$. For any $\chi(x)$ that satisfies Eq. \eqref{tH} \cite{Weller:2021wog}:
\begin{equation}
    \int \mathrm{d} x  \; \left( \mathcal{P} \int_0^1 {\rm d}y \frac{\chi(y)}{(x-y)^2} \right)  = 0\;.
\label{constraint-tH}
\end{equation}
After integrating Eq. \eqref{tH} using Eq. \eqref{constraint-tH} we find
\begin{equation}
	\frac{M_\pi^2}{m_q^2}=\frac{2}{\beta} + 4 \;.
\label{identity-tH}
\end{equation}
For $\beta \ll 1$, Eq.~\eqref{end-points-tH-A} implies that $\beta=m_q/(\sqrt{3} g \pi)$, so that Eq.~\eqref{identity-tH} implies $M^2_\pi \propto m_q$ to leading order in $m_q$, i.e. we recover the GMOR relation. Numerical solution also confirms that the 't Hooft equation is consistent with the GMOR relation \cite{PhysRevD.107.034004}.
Eq.~\eqref{tH} therefore predicts that $\chi(x) \to 1$ in the chiral limit. 

\begin{table}
\centering
\begin{tabularx}{0.45\textwidth}{|>{\centering\arraybackslash}X|>{\centering\arraybackslash}X|>{\centering\arraybackslash}X|>{\centering\arraybackslash}X|}
\hline
         & $\kappa$/MeV & $m_q$/MeV & $g$/MeV \\ \hline
    tH-A & 423 & 59.3 & 30.2\\ \hline
    tH-B & 423 & 55.3 & 41.2\\ \hline
    tH-C & 423 & 47.2 & 66.5 \\ \hline
\end{tabularx}
    \caption{Extracted values of the parameters of the tH models discussed in the text.}
    \label{tab:tH}
\end{table}

\begin{figure}[ht]
\centering 
\includegraphics[width=12.5cm]{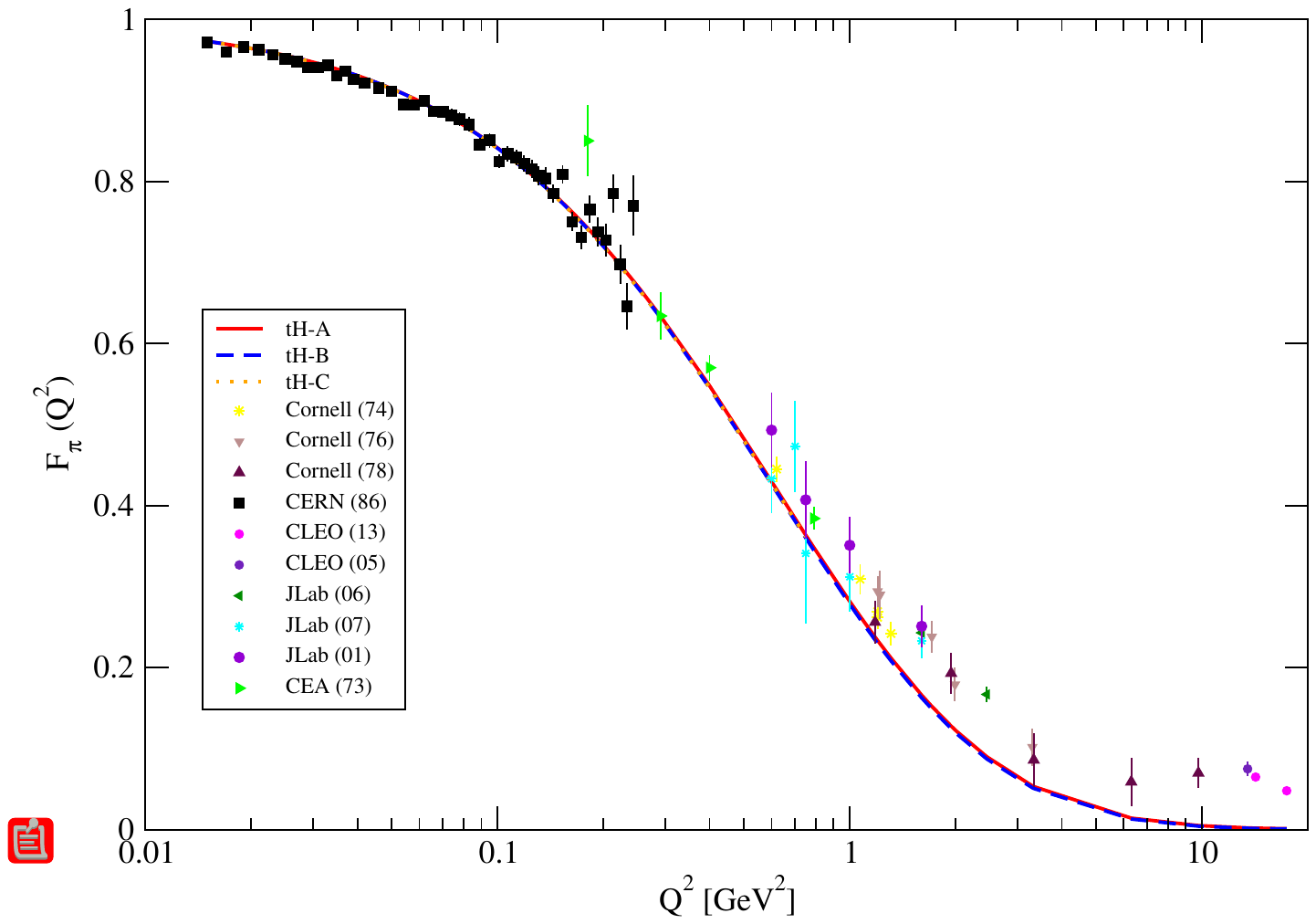}\\ \includegraphics[width=12.5cm]{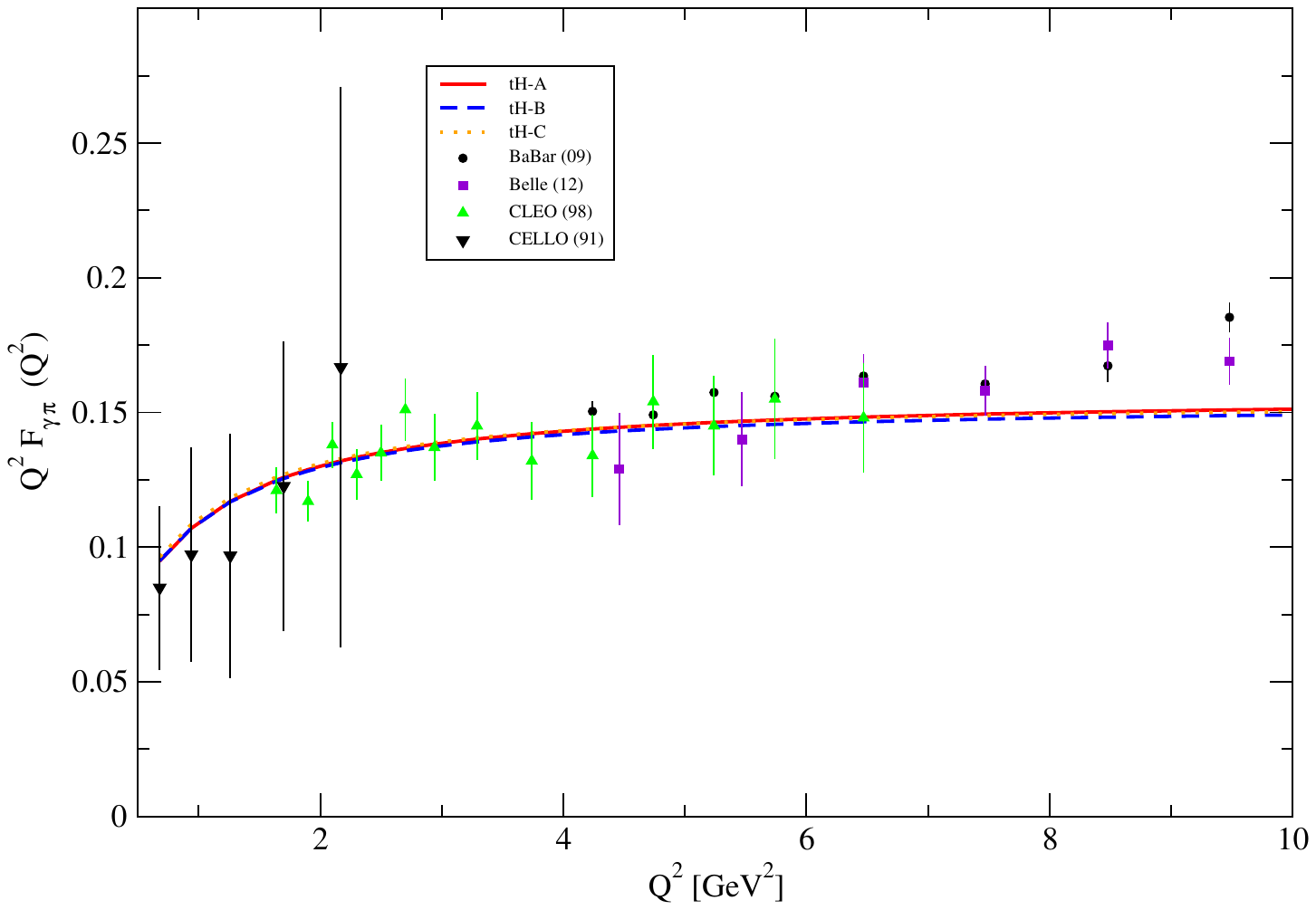}\caption{Postdictions of the tH models for the FF data \cite{Amendolia:1986wj,Bebek:1974,Bebek:1976,Bebek:1978,Volmer:2001,Horn:2006tm,Pedlar:2005sj,Seth:2012nn} and TFF data \cite{Uehara2012,Aubert2009,Gronberg1998, 
Behrend1991}.}
\label{Fig:FF-tH}
\end{figure}

Following the same methodology as in the previous section, we show the result of fixing the tH model parameters using the measured values of the charged pion mass, decay constant and radius in Table \ref{tab:tH}. Our postdictions for the FF data using the tH model are shown in Fig. \ref{Fig:FF-tH} and, not surprisingly, the results are almost identical to those in the LV model.

\clearpage

Using the $Q^2 \to 0$ limit of Eq.~\eqref{TFF}, we obtain $\Gamma_{\gamma\gamma} = 7.2, 7.4, 7.6$~eV for tH-A, tH-B, tH-C, which is very similar to the the corresponding numbers for the LV models.

\section{A third model}
\label{Sec:FS}
To investigate the correlation between the tH and LV models we introduce a third model:
\begin{equation}
\left(\frac{m^2_q-g^2}{x(1-x)}\right)\chi(x) - g^2 \mathcal{P} \int_0^1 {\rm d}y \frac{\chi(y)}{(x-y)^2}-\sigma^2 \partial_x (x(1-x) \partial_x) \chi(x)=M_\parallel^2 \chi(x) ,
  \label{FSA}
\end{equation}
which we refer to as the FS-A model.  If $m_q=g$, $4g_s=g^2/\sigma^2$, $\mu^2=M_\parallel^2/g^2$ and $x=z$, Eq. \eqref{FSA} becomes the quantum spectral curve of a four-segmented string in $\mathrm{AdS}_3$ derived by Vegh \cite{Vegh:2023snc}:  
\begin{equation}
-\mathcal{P}\int_0^1 {\rm d}z^\prime \frac{\chi(z^\prime)}{(z-z^\prime)^2}-\frac{1}{4g_s}\partial_z (z(1-z) \partial_z) \chi(z)=\mu^2  \chi(z) \;,
	\label{AdS1}
\end{equation}
where $g_s$ is the string tension in units of the AdS radius (squared).

If instead we choose  $m_q^2=g^2+\sigma^2/4$, Eq. \eqref{FSA} becomes 
\begin{equation}
-\mathcal{P}\int_0^1 {\rm d}z^\prime \frac{\chi(z^\prime)}{(z-z^\prime)^2}-\frac{1}{4g_s}  \sqrt{z(1-z)} \partial^2_z (\sqrt{z(1-z)} \chi(z))=\mu^2  \chi(z) \;,
	\label{AdS2}
\end{equation}
where we have used the fact that
\begin{equation}
\partial_x (x(1-x) \partial_x) \chi(x)= \sqrt{x(1-x)} \partial_x^2 (\sqrt{x(1-x)} \chi(x)) + \frac{1}{4x(1-x)} \chi(x)\;.
\end{equation}
Eq. \eqref{AdS2} is a second possibility for the string equation in $\mathrm{AdS}_3$. The two possibilities arise from the two ways to symmetrize the weak-coupling term in the string Hamiltonian \cite{Vegh:2023snc}. Notice that the constraint, $m_q^2 = g^2 + \sigma^2/4$, together with $4g_s=g^2/\sigma^2$ implies that $g\le m_q$ since $g_s \ge 0$.
 
\begin{figure}[htbp]
\centering 
\includegraphics[width=13cm]{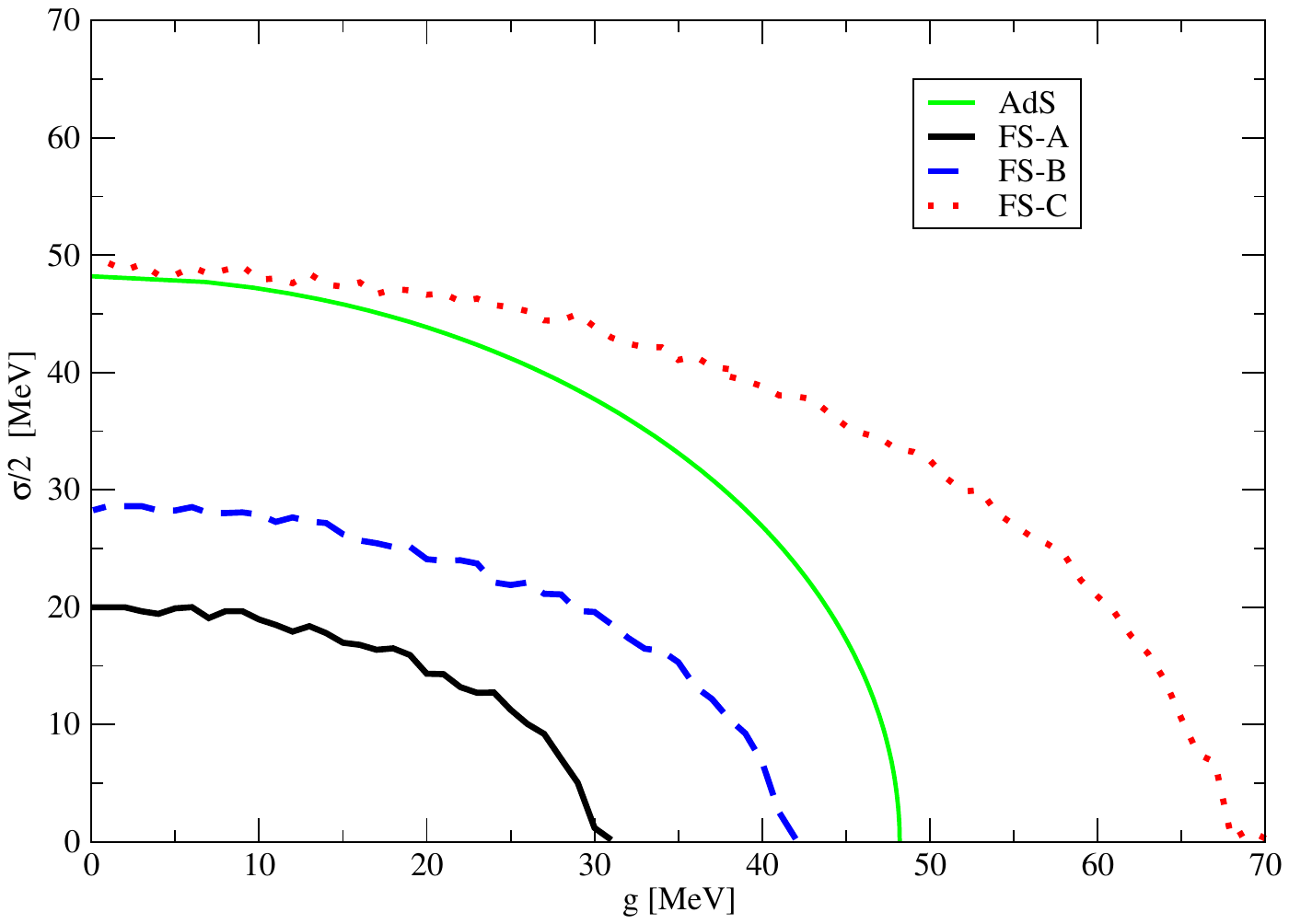} 
\caption{Exploring the correlation between the tH and LV potentials via the FS models. Each point on a given curve represents a pair $(g,\sigma)$ of extracted parameters using the pion data. Note that $m_q$ varies very weakly along each curve between its tH and LV limits: see Tables \ref{tab:LV} and \ref{tab:tH}. The solid green curve, labelled as AdS, is the holographic constraint $g^2 + \sigma^2/4=m_q^2$ with $m_q=48.2$~MeV (which lies in between its tH-C and LV-C limits).}
\label{Fig:contours}
\end{figure}  

Fig.~\ref{Fig:contours} explore the correlation between the tH and LV models. The quark mass varies very weakly along each ellipse. Notice that FS-C is able to accommodate the holographic constraint, $g^2 + \sigma^2/4=m_q^2$ in  the $\sigma \gg g$ limit, which is equivalent to a weak string coupling, $g_s \ll 1$. Viewed this way, one could regard our observation that $m_q = \sigma/2$ in the LV-C model as a prediction of the AdS$_3$ string equation. Though without a firmer basis for the correspondence this is speculative.

Finally, we should comment upon the fact that our extracted value of $\kappa=423$ MeV is much lower than the typical value, $\kappa \approx 500$ MeV, that fits the Regge slopes for light mesons \cite{Brodsky:2014yha}. Previous work on the pion using holography also require smaller values of $\kappa$: $\kappa=375$~MeV in \cite{Brodsky:2007hb}, $\kappa=361$~MeV in \cite{PhysRevD.79.055003}, $\kappa \approx 370$~MeV in \cite{Bacchetta:2017vzh}, and $\kappa=432$~MeV in  \cite{Brodsky:2011xx}. 

\section{Conclusions}
We have used the 't Hooft and Li-Vary potentials to model the longitudinal dynamics in the pion. For the transverse dynamics we use the well-established light-front holographic Schr\"odinger equation. Using only the low energy pion data to fix the parameters of the models, we find very good agreement with the low $Q^2$ form factor data, and a longitudinal momentum distribution that is considerably more peaked about $x \sim 1/2$ than in previous studies. We have also explored the degeneracy between the tH and LV models. Our analysis of the longitudinal dynamics is in accord with an equation derived by Vegh to describe the dynamics of a four-segmented string in AdS$_3$. 

\appendix

\section{Derivation of the Transition Form Factor}
\label{App:TFF}

\begin{figure}[ht]
\centering 
\includegraphics[width=6cm]{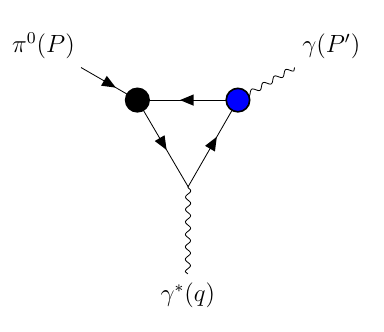}
\caption{Loop diagram for the TFF. The black blob represents the pion as a QCD bound state, encoded in the pion light-front wavefunction. The blue blob represents the hadronic fluctuation of the real photon encoded in the photon light-front wavefunction. The point-like coupling of the virtual photon represents the EM current through which the pion-to-photon transition occurs. There is a second diagram, with the arrows in the loop reversed.}
\end{figure}

The pion-to-photon transition form factor, $F_{\pi \gamma}(Q^2)$, defined via \cite{PhysRevD.102.036005}
\begin{equation}
	\langle \gamma (P^\prime) | J_{\mathrm{EM}}^\mu | \pi^0 (P) \rangle= i F_{\pi \gamma}(Q^2) \epsilon^{\mu \nu \rho \sigma} P_\nu \varepsilon_\rho q_\sigma 
\label{TFF-def}
\end{equation}
where $J_{\mathrm{EM}}^\mu$ is the quark EM current given by
\begin{equation}
	J_{\mathrm{EM}}^\mu= e_f \bar{\Psi}(0) \gamma^\mu \Psi(0) 
\end{equation}
with $e_f$ being the electric charge, in units of $e$, of the quark (flavour $f$) and $(\bar{\Psi}) \Psi$ are the (adjoint) quark-field operators evaluated at the same spacetime point. The $4$-momentum of  the virtual photon is $q_\sigma=P_\sigma^\prime - P_\sigma$,  with $q^2\equiv -Q^2$, while $\epsilon^{\mu \nu \rho \sigma}$ is the Levi-Civita totally antisymmetric tensor and $\varepsilon_\rho$ is polarization $4$-vector of the transversely-polarized real photon. In the $q^+=0$ frame, 
\begin{equation}
    P^\mu=\left(P^+, \frac{M_\pi^2+Q^2}{P^+}, \mathbf{q}\right) \;,
\label{P}
\end{equation}
and
\begin{equation}
    q^\mu=\left(0, -\frac{M_\pi^2+Q^2}{P^+}, -\mathbf{q}\right)
\label{q}
\end{equation}
so that $Q^2=q_\perp^2$ where $q_\perp=|\mathbf{q}|$. For the polarization vector of the real photon, we choose
\begin{equation}
    \varepsilon_\rho=\frac{1}{\sqrt{2}} (0,0,1, i) \;.
\label{epsilon}
\end{equation}
 Taking $\mu=+$ in Eq.~\eqref{TFF-def}, and using Eqs.~\eqref{P}, \eqref{q} and \eqref{epsilon}, yields
\begin{equation}
	\langle \gamma (P^\prime) | J_{\mathrm{EM}}^+ | \pi^0 (P) \rangle=\frac{1}{\sqrt{2}} F_{\pi \gamma}(Q^2) P^+ Q \;.
\label{TFF-def-simple}
\end{equation}
The Fock expansion of  the matrix element on the left-hand-side is
\begin{align}
	\langle \gamma (P^\prime) | J_{\mathrm{EM}}^\mu | \pi^0 (P) \rangle =&  e_f \sqrt{4 \pi N_c} \sum_{h,\hbar} \int \frac{\mathrm{d} x  \;\mathrm{d}^2 \mathbf{k}}{16 \pi^3} \Psi_{h,\bar{h}}^{*\gamma} (x, \mathbf{k}-(1-x) \mathbf{q}) \;\Psi^{\pi}_{h,\bar{h}} (x, \mathbf{k}) \nonumber \\ & \times \left\{ \frac{\bar{v}_{\bar{h}}(1-x,-\mathbf{k})}{\sqrt{x(1-x)}} \gamma^\mu \frac{u_h(x,\mathbf{k})}{\sqrt{x}} \right\},
\label{Fock-expansion}
\end{align}
where $h(\bar{h})=\pm$ are the quark(antiquark) helicities. The factor $\sqrt{N_c}$ originates from the product of two colour factors: $N_c$ from the sum over all colours in the quark loop and $1/\sqrt{N_c}$ from the colour singlet wavefunction for the pion. The factor $\sqrt{4\pi}$ accounts for a mismatch of convention between the normalization of the meson wavefunction used in this paper and that used in the light-front formalism. Specifically, we use \cite{Forshaw:2003ki}
\begin{equation}
    \sum_{h,\bar{h}}\int \frac{\mathrm{d}^2\mathbf{k}}{4\pi^2} \mathrm{d} x |\Psi_{h,\bar{h}}(x,\mathbf{k})|^2 =1,
\end{equation}
which is consistent with  Eq. \eqref{normalization-b}, whereas the light-front formalism uses \cite{Lepage:1980fj}
\begin{equation}
   \sum_{h,\bar{h}} \int \frac{\mathrm{d}^2\mathbf{k}}{16\pi^3} \mathrm{d} x |\Psi_{h,\bar{h}}(x,\mathbf{k})|^2 =1 \;.
\end{equation}
In Eq. \eqref{Fock-expansion}, $\Psi_{h,-\hbar}^{* \gamma}(x, \mathbf{k}-(1-x) \mathbf{q})$ is the  real photon light-front wavefunction (complex-conjugated) and  $\Psi^\pi_{h,\hbar}(x, \mathbf{k})$ is the pion light-front wavefunction. 
Now, taking $\mu=+$ in Eq.~\eqref{Fock-expansion}, we find
\begin{equation}
	\langle \gamma (P^\prime) | J_{\mathrm{EM}}^+ | \pi^0 (P) \rangle= 2 P^+ e_f \sqrt{4 \pi N_c} \sum_{h,\hbar} \int \frac{\mathrm{d} x \mathrm{d}^2 \mathbf{k}}{16 \pi^3} \Psi_{h,\bar{h}}^{*\gamma} (x, \mathbf{k}-\bar{x} \mathbf{q}) \Psi^{\pi}_{h,\bar{h}} (x, \mathbf{k}) \delta_{h,-\bar{h}} \;,
\end{equation}
where we have used the fact that \cite{Lepage:1980fj}
\begin{equation}
\frac{\bar{v}_{\bar{h}}(1-x,-\mathbf{k})}{\sqrt{x(1-x)}} \gamma^+ \frac{u_h(x,\mathbf{k})}{\sqrt{x}} = 2P^+ \delta_{h,-\bar{h}} \;.
\label{plus-current}
\end{equation}
Fourier transforming to $\mathbf{b}$-space, we obtain
\begin{equation}
 	\langle \gamma (P^\prime) | J_{\mathrm{EM}}^+ | \pi^0 (P) \rangle=2P^+ e_f \sqrt{\frac{N_c}{4\pi}} \int \mathrm{d} x \, \mathrm{d}^2 \mathbf{b} \; e^{-i \bar{x}\mathbf{q} \cdot \mathbf{b}} \sum_{h,\bar{h}} \Psi_{h,\bar{h}}^{*\gamma} (x, \mathbf{b}) \Psi^{\pi}_{h,\bar{h}} (x, \mathbf{b}) \delta_{h,-\bar{h}},
 \label{bspace}
\end{equation}
where the real photon wavefunction, derived using perturbative QED, is given by \cite{Cox:2009ag}
\begin{equation}
	\Psi^\gamma_{h,\hbar}(x,\mathbf{b})= -\sqrt{2} e_f \left\{ i  e^{i \varphi} (x \delta_{h+,\bar{h}{-}}- (1-x)\delta_{h-,\bar{h}{+}} ) \frac{m_q K_1(m_q b_\perp)}{2\pi} + \delta_{h+, \bar{h}+} \frac{K_0(m_q b_\perp)}{2\pi} \right\} \;.
\label{photon-b-space}
\end{equation}
Note that Eq. \eqref{photon-b-space} differs from the photon wavefunction given in \cite{Cox:2009ag} by an overall  factor of $\sqrt{N_c/(4\pi)}$ on the right-hand-side. This is because we have already extracted the overall colour factor in Eq. \eqref{Fock-expansion} and we 
have implicitly multiplied the photon wavefunction by $\sqrt{4\pi}$ to account for the normalization-mismatch mentioned above. The pion wavefunction is given by  
\begin{equation}
	\Psi_{h,\hbar}^\pi (x,\mathbf{b}) = \frac{\kappa}{\sqrt{\pi}} X(x)  \exp \left(-\frac{\kappa^2 x(1-x) b_\perp^2}{2}\right) \frac{1}{\sqrt{2}} h \delta_{h,-\hbar} \;,
 \label{pion-b-space}
\end{equation}
i.e. the product of Eq. \eqref{pionwf-gaussian} with the pion's helicity wavefunction.\footnote{We assume that this helicity wavefunction is momentum-independent. For momentum-dependent helicity wavefunctions, see \cite{Ahmady:2016ujw,Choi:2020xsr,Chang:2016ouf}.} 

Inserting Eqs.~\eqref{photon-b-space} and \eqref{pion-b-space} in Eq.~\eqref{bspace}, summing over all helicities, and performing the angular integration, we find
\begin{align}
\langle \gamma (P^\prime) | J_{\mathrm{EM}}^+ | \pi^0 (P) \rangle &= P^+ e_f^2 \kappa \frac{\sqrt{N_c}}{\pi} \int \mathrm{d} x X(x) \int \mathrm{d} b_\perp m_q b_\perp K_1(m_q b_\perp) \nonumber \\ & \times \exp \left(-\frac{\kappa^2x(1-x)b_\perp^2}{2} \right) J_1(b_\perp (1-x) Q).
\label{after-angular-int}
\end{align}
Since the flavour wavefunction of the neutral pion is $|\pi^0\rangle= \frac{1}{\sqrt{2}} | u\bar{u} - d\bar{d} \rangle$, we must take $e^2_f= \frac{1}{\sqrt{2}} (e_u^2 - e_d^2) = \frac{1}{3\sqrt{2}}$. Since, $e_{\bar{f}}^2=e_f^2$, the second Feynman diagram, in which the fermion line in the triangle loop are reversed, give exactly the same contribution. Taking both diagrams into account (i.e. multiplying by right-hand-side of Eq. \eqref{after-angular-int} by $2$), and inserting Eq. \eqref{after-angular-int} in Eq. \eqref{TFF-def-simple}, we find
\begin{eqnarray}
\label{TFF-Appendix}
Q^2 F_{\pi \gamma} (Q^2)&=&\frac{2 \kappa Q}{\sqrt{3} \pi}  \int_0^1 \mathrm{d} x X(x)
	\int_0^\infty \mathrm{d} b_\perp (m_q b_\perp) K_1(m_q b_\perp) \\ \nonumber 
	&\times& \exp\left(-\frac{\kappa^2 x(1-x) b_\perp^2}{2} \right) 
	J_1(b_\perp (1-x) Q),
    \end{eqnarray}
which is  Eq. \eqref{TFF} given that $X(x)=\sqrt{x(1-x)}\chi(x)$. 

\section{Chiral limit behaviour}
\label{App:chiral}
In the chiral limit, models A, B and C predict that $\chi^{(0)}(x)=1$, $\chi^{(0)}(x)=\sqrt{6x(1-x)}$ and  $\chi^{(0)}(x)=\sqrt{30}x(1-x)$ respectively\footnote{The superscript indicates we take the massless quark limit.}. Model A thus predicts that $X^{(0)}(x)=\sqrt{x(1-x)}$ which is often referred to as the longitudinal mode of light-front holography. More precisely, $X(x)=\sqrt{x(1-x)}$ results from the holographic mapping of the pion electromagnetic or gravitational form factor in physical spacetime and $\mathrm{AdS}_5$ with a hard-wall geometry \cite{Brodsky:2006uqa,Brodsky:2007hb,Brodsky:2008pf} which is different from the $\mathrm{AdS}_5$ soft-wall geometry that generates Eq.~\eqref{holographic-potential}.

We now investigate the extent to which the chiral limit of our models satisfy two important results: the Brodsky-Lepage (BL) limit and the Adler-Bell-Jackiw (ABJ) chiral anomaly relation. The first result concerns the  celebrated Brodsky-Lepage  formula for the transition form factor \cite{Lepage:1980fj}:
\begin{equation}
	 Q^2 F_{\pi \gamma} (Q^2) = \frac{\sqrt{2}f_\pi}{3} \int \frac{\mathrm{d} x}{(1-x)} \varphi_\pi(x, (1-x)Q) + \mathcal{O} \left(\alpha_s, \frac{m_q^2}{Q^2} \right)\;, 
\label{BL}
\end{equation}
where $\varphi_\pi(x,(1-x)Q)$ is the pion Distribution Amplitude (DA). Taking the  $Q^2 \to \infty$ and $m_q \to 0$ limit, Eq.~\eqref{BL} becomes
\begin{equation}
	\lim_{Q^2 \to \infty} Q^2 F^{(0)}_{\pi \gamma} (Q^2) = \frac{\sqrt{2}f^{(0)}_\pi}{3} \int \frac{\mathrm{d} x}{(1-x)} \varphi_\pi(x, \infty),
\label{BL-chiral-largeQ2}
\end{equation}
where  $\varphi_\pi(x, \infty)$ is the asymptotic DA given by
\begin{equation}
	\varphi_\pi(x, \infty)=6 x(1-x) ,
\label{asymp}
\end{equation}
as obtained in conformal QCD \cite{Braun:2003rp}.  Substituting Eq. \eqref{asymp} in Eq. \eqref{BL-chiral-largeQ2} results in
\begin{equation}
    \lim_{Q^2 \to \infty} \frac{Q^2 F^{(0)}_{\pi \gamma} (Q^2)}{\sqrt{2} f_\pi^{(0)}}=1 \;,
\label{BL-F-fpi}
\end{equation}
which is the BL limit. The second result is the ABJ relation, Eq.~\eqref{ABJ-lit}, which we rewrite as
\begin{equation}
 2\sqrt{2} \pi^2	F^{(0)}_{\pi \gamma}(0)f_\pi^{(0)} = 1 
 \label{ABJ}
 \end{equation}
 to emphasize that the ABJ relation is derived in the chiral limit.\footnote{A pedagogical derivation of Eq.~\eqref{ABJ} can be found in \cite{Khodjamirian:2020btr}.}

We now proceed to find the chiral limits of Eq.~\eqref{fpi} and  Eq.~\eqref{TFF},  the latter both in the $Q^2 \to \infty$ and $Q^2 \to 0$ limits. The chiral limit of Eq.~\eqref{fpi} is 
\begin{equation}
	f_\pi^{(0)}=\frac{\sqrt{6}}{\pi} \kappa \int \mathrm{d} x \sqrt{x(1-x)} \chi^{(0)}(x) \;.
\label{fpi-chiral}
\end{equation}
To find the chiral limit of Eq.~\eqref{TFF-Appendix}, note that when $m_q \to 0$,  $(m_q b_\perp) K_1(m_q b_\perp) \to 1$, since the Gaussian in Eq.~\eqref{TFF-Appendix} 
exponentially suppresses large $b_\perp$ for any reasonable $\chi^{(0)}(x)$. The $b_\perp$ integral in Eq. \eqref{TFF} can then be performed, yielding
\begin{equation}
	Q^2 F^{(0)}_{\pi \gamma} (Q^2)=\frac{2 \kappa}{\sqrt{3} \pi}  \int \mathrm{d} x \sqrt{x(1-x)} \chi^{(0)}(x) \left(1- \exp\left(-\frac{(1-x)Q^2}{2\kappa^2x} \right) \right) \frac{1}{(1-x)} \;.\label{TFF-chiral}
\end{equation}
If $Q^2 \to \infty$, Eq. \eqref{TFF-chiral} becomes
\begin{equation}
\lim_{Q^2 \to \infty} Q^2 F^{(0)}_{\pi \gamma} (Q^2)=\frac{2 \kappa}{\sqrt{3} \pi}  \int \mathrm{d} x \sqrt{\frac{x}{(1-x)}} \chi^{(0)}(x) \;.
	\label{TFF-chiral-large-Q2}
\end{equation}
On the other hand, taking $Q^2 \to 0$ limit of Eq. \eqref{TFF-chiral} we obtain: 
\begin{equation}
F^{(0)}_{\pi \gamma}(0)= \frac{1}{\sqrt{3} \pi \kappa} \int \mathrm{d} x \sqrt{\frac{(1-x)}{x}} \chi^{(0)}(x) \;. 
	\label{TFF-chiral-lowQ2}
	\end{equation}
Notice that any model with $n<0$ leads to end-point divergences.  

\begin{table}
\centering
\begin{tabularx}{0.75\textwidth}{|>{\centering\arraybackslash}X|>{\centering\arraybackslash}X|>{\centering\arraybackslash}X|}
\hline
      & $2 \sqrt{2} \pi^2 F^{(0)}_{\pi \gamma}(0)f_\pi^{(0)}$ & $\lim_{Q^2 \to \infty} \frac{Q^2 F^{(0)}_{\pi \gamma} (Q^2)}{\sqrt{2} f_\pi^{(0)}}$ \\ \hline
Model A  & $2.46$& $1.33$\\ \hline
 Model B & $2$& $1$\\ \hline  
 Model C  &$1.74$&$0.89$ \\ \hline    
     \end{tabularx}
    \caption{Comparing to the ABJ and BL predictions.}    
    \label{tab:ABJ-BL}
\end{table}

Table \ref{tab:ABJ-BL} indicates the extent to which Models A -- C accord with ABJ and BL. The BL limit is satisfied exactly for Model B and approximately for Models A and C. On the other hand, none of the models satisfy the ABJ relation. Model B overestimates the ABJ prediction by exactly a factor of $2$. This result was reported previously in \cite{Brodsky:2011xx,Musatov:1997pu}. In fact, Eq. \eqref{TFF-chiral} with Model B for $\chi^{(0)}(x)$, coincides with holographic TFF derived in \cite{Brodsky:2011xx}.\footnote{More precisely, \cite{Brodsky:2011xx} replaces the right-hand-side of Eq. \eqref{normalization-b} by $P_{q\bar{q}}$. If we take $P_{q\bar{q}}=1$, then the holographic TFF of \cite{Brodsky:2011xx} coincides exactly with Eq. \eqref{TFF-chiral} (with $\chi^{(0)}(x)$ given by Model B) and also with the TFF derived in \cite{Musatov:1997pu}.} This should all be set in the context of the fact that our results with $m_q \neq 0$ do agree with the ABJ relation.

\begin{figure}[ht]
\centering 
\includegraphics[width=12cm]{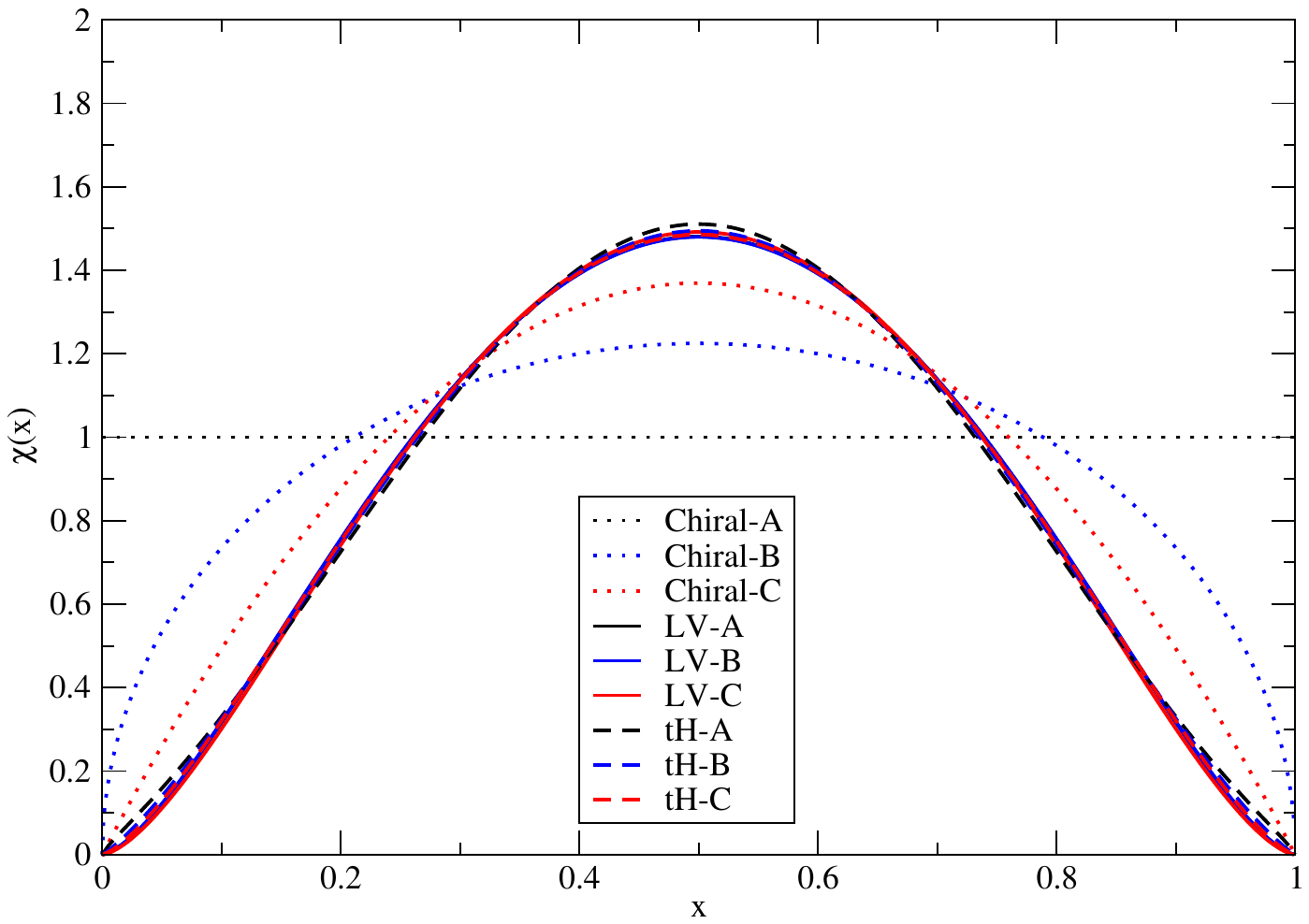}\caption{The extracted (and degenerate) longitudinal wavefunction, $\chi(x)$, compared to its chiral limit form in the three models. }
\label{Fig:chi}
\end{figure}

Finally, we illustrate the degree of chiral symmetry breaking in the models in Fig. \ref{Fig:chi}. Model C deviates least from its chiral form. 

\section{Acknowledgements}
JRF thanks Acadia University for the award of a Harrison McCain Foundation Visiting Professorship. RS is supported by an Individual Discovery Grant (SAPIN-2020-00051) from the Natural Sciences and Engineering Research Council of Canada (NSERC) and thanks the Particle Physics Group at the University of Manchester for their hospitality. We thank David Vegh and Yang Li for clarifying remarks on their work.

\bibliographystyle{apsrev4-2}
\bibliography{paper.bib}    
\end{document}